\documentclass{ifacconf}

\usepackage{natbib}        

\usepackage{graphics, graphicx}
\usepackage{epsfig, enumerate}
\usepackage{amsmath, amssymb, amsfonts}
\usepackage{bm}
\usepackage{times, stmaryrd}
\usepackage{newtxtext} 
\usepackage{array, multirow, dsfont}
\usepackage{subeqnarray, stmaryrd, mathtools, bm}
\usepackage{booktabs}
\usepackage{threeparttable}
\usepackage{xcolor}
\usepackage{bbm}
\usepackage{subfigure} 

\usepackage[ruled, vlined, linesnumbered,algo2e]{algorithm2e}

\newtheorem{remark}{Remark}
\newtheorem{definition}{Definition}

\DeclareMathOperator{\F}{\mathrm{F}}
\DeclareMathOperator{\G}{\mathrm{G}}
\DeclareMathOperator{\U}{\mathrm{U}}
\DeclareMathOperator{\I}{\mathrm{I}}
\begin{document}
\begin{frontmatter}

\title{Optimal Planning and Control under Signal Temporal Logic Specifications\thanksref{footnoteinfo}} 

\thanks[footnoteinfo]{This work was supported by the Fundamental Research Funds for the Central Universities under Grant DUT22RT(3)090, the LiaoNing Revitalization Talents Program (XLYC2403048), and the National Natural Science Foundation of China under Grants 62573085.}

\author[First]{Zuodong Pan}, 
\author[First]{Xu Fang}, 
\author[First]{Wei Ren}
 

\address[First]{Key Laboratory of Intelligent Control and Optimization for Industrial Equipment of Ministry of Education, Dalian University of Technology, Dalian 116024, China. (zuodongpan@mail.dlut.edu.cn, xufang@dlut.edu.cn, wei.ren@dlut.edu.cn)}

\begin{abstract}                
This paper addresses the planning and control problem for nonlinear systems under Signal Temporal Logic (STL) specifications. We first decompose an STL task into finite local tasks. A sampling-based method generates sequences of local waypoints to satisfy all local tasks, from which the corresponding satisfaction pair sets are derived. Following a local-to-global strategy, all sequences of local waypoints are synthesized into a global one, based on which a safe corridor is then constructed. Leveraging the safe corridor and the satisfaction pair sets, an optimization problem is formulated and solved to derive a position trajectory that satisfies the STL task. Finally, numerical examples and comparative results are presented to demonstrate the efficacy of the proposed approach. 
\end{abstract}


\end{frontmatter}
\section{Introduction}
\label{sec-intro}

In planning and control applications such as mobile robots, recent research has shifted from simple tasks to complex ones \citep{plaku2015motion}. Simple tasks involve the consensus, stability and tracking, while complex tasks include the sequencing, coverage, recurrence, and reachability \citep{fainekos2009temporal}. In order to describe complex tasks, many concise and expressive languages have been proposed and extensively applied, such as linear temporal logic (LTL) and signal temporal logic (STL) \citep{baier2008principles}. LTL provides a Boolean satisfaction of properties by given signals, and the control design for LTL tasks usually requires discrete abstractions of dynamical systems and is based on time-consuming graph search methods \citep{belta2007symbolic, ren2024zonotope, fainekos2009temporal}. Since LTL does not involve explicit time constraints, STL is proposed to embed explicit time intervals into the task specifications on the one hand and to provide quantitative semantics for the evaluation of the task satisfaction over a continuous-time signal on the other hand, thereby avoiding discrete abstractions of dynamical systems. This extension enables the definition of the completion time intervals and the assessment of the robustness in the task satisfaction \citep{yao2025model}.

An STL formula is non-nested or nested, depending on whether it contains more than one temporal operator. Non-nested formulas involve sequentially reaching the regions of interest while avoiding the obstacles. Regarding dynamical systems under non-nested formulas, many control methods have been developed, including, to name a few, sampling-based approaches \citep{linard2023real, karlsson2020sampling}, control barrier function (CBF)-based methods \citep{lindemann2019control, garg2019control}, and task decomposition methods \citep{ren2022reachability, tan2024decomposition}. While these methods can handle most non-nested formulas, they struggle with nested formulas, which involve remaining within, and revisiting regions of interest.

To address systems with a mixture of non-nested and nested formulas, there are some endeavors in recent years. In the realm of planning, one approach encodes the STL specifications into a mixed-integer linear programming (MILP) problem to derive a trajectory \citep{sun2022multi}. Another planning method generates a satisfying trajectory by iteratively sampling in the spatio-temporal (ST) space \citep{sewlia2023cooperative}. Separately, for control synthesis, a CBF-based method is proposed in \citep{yu2024continuous}. The method introduces a notion of STL tree, which is used to design CBFs and update activation time intervals online. The control signal is given by an online CBF-based program. However, the application of sampling-based methods to address the mixed non-nested and nested formulas remains an open challenge, and is the primary focus of this work.

This paper proposes a planning and control approach to address an STL task for nonlinear systems. In the planning stage, inspired by \citep{ren2022reachability} and \citep{sewlia2023cooperative}, we decompose the STL task into finite local tasks, each of which can be satisfied by a sequence of local waypoints derived via a sampling-based method. Specifically, the union application time interval (UATI) of the STL task is divided into multiple local UATIs. Within each local UATI, all local sub-tasks are identified and combined using conjunctions to form a local task. For each local task, a local path is derived by iteratively sampling in the ST-space. The path is subsequently discretized into a sequence of points via linear interpolation. If these points satisfy the local task, they are regarded as the sequence of local waypoints, and the corresponding satisfaction pair set is derived; otherwise, the above process is repeated until the local task is satisfied. Finally, a local-to-global strategy is applied to synthesize all sequences of local waypoints into a global one that satisfies the STL task. In the control stage, motivated by \citep{9293348}, we first construct a safe corridor via the sequence of global waypoints. Then, based on the safe corridor and the satisfaction pair sets, an optimization problem is formulated and solved to derive the optimal control strategy such that the corresponding position trajectory satisfies the STL task.

The rest of this paper is organized as follows. Preliminaries and problem formulation are stated in Section \ref{sec-PreAndPro}. Section \ref{sec-Methodology} presents the proposed planning and control approach. Simulation results are given in Section \ref{sec-Simulation}. Finally, Section \ref{sec-Conclusion} concludes this paper. 

\section{Preliminaries and Problem Formulation}
\label{sec-PreAndPro}

Let $\mathbb{R} := (-\infty, +\infty)$, $\mathbb{R}_{\ge 0}  := [0, +\infty)$, $\mathbb{N}_{\ge 0}  := \{0, 1, \ldots\}$, $\mathbb{N}_{> 0}  := \{1, 2, \ldots\}$ and $\mathbb{R}^{n}$ be the $n$-dimensional Euclidean space. Given a finite set $\mathbf{A}\subset\mathbb{R}^{n}$, $|\mathbf{A}|$ denotes the cardinality of $\mathbf{A}$. For $\mathbf{A}, \mathbf{B} \subset \mathbb{R}^{n}$, $\mathbf{B}\setminus\mathbf{A} := \{x: x\in\mathbf{B}, x\notin\mathbf{A}\}$. Given a compact interval $\mathbf{C} := [a, b]$ with $a, b \in \mathbb{R}_{\ge 0} $, we denote $\underline{\mathbf{C}} := a$ and $\overline{\mathbf{C}} :=  b$. The interior of $\mathbf{C}$ is denoted by $\mathbf{C}^{\circ} := (a, b)$. For two compact intervals $\mathbf{C} = [a, b]$ and $\mathbf{D} = [c, d]$ with $a, b, c, d \in \mathbb{R}_{\ge 0} $, the Minkowski sum is given by $\mathbf{C} \oplus \mathbf{D} := [a+c, b+d]$. For a finite set $\mathbf{S} \subset \mathbb{R}$ with $|\mathbf{S}| = N+1$, its ordered sequence is denoted by $\mathtt{ord}(\mathbf{S}) := (\xi_i)_{i=0}^{N}$, where $\xi_i \in \mathbf{S}$ and $\xi_0 < \xi_1 < \dots < \xi_N$. Given a conjunctive normal form $\Psi = \wedge_{i=1}^N \psi_i$, its clause set is expressed as $\mathtt{cla}(\Psi) := \{\psi_i \}_{i=1}^N$. For a point $\boldsymbol{x}:= (x_1, \dots, x_{n+m}) \in \mathbb{R}^{n+m}$, and two compact sets $\mathbf{A} \subset \mathbb{R}^{n}$ and $\mathbf{E} \subset \mathbb{R}^{m}$, the projections of $\boldsymbol{x}$ onto $\mathbf{A}$ and $\mathbf{E}$ are denoted by $\mathtt{proj}(\boldsymbol{x}, \mathbf{A}) := (x_1, \dots, x_n)$ and $\mathtt{proj}(\boldsymbol{x}, \mathbf{E}):= (x_{n+1}, \dots, x_{n+m})$, respectively.

\subsection{Signal Temporal Logic}
\label{subsec-STL}

Let $\boldsymbol{x} : \mathbb{R}_{\ge 0}  \rightarrow \mathbb{R}^{n}$ be a continuous-time signal. The predicate $\mu$ at time $t \ge 0$ is obtained after the evaluation of a predicate function $g_{\mu} : \mathbb{R}^{n} \times \mathbb{R}_{\ge 0}  \rightarrow \mathbb{R}$. That is,   
\begin{equation}
\mu:=
  \begin{cases}
    \top, & \mathrm{if}\quad g_\mu(\boldsymbol{x}(t))\geq0, \\
    \bot, & \mathrm{if}\quad g_\mu(\boldsymbol{x}(t))<0,
  \end{cases}
\end{equation}
where $\top$ is true and $\bot$ is false. Signal Temporal Logic (STL) \citep{maler2004monitoring} is a predicate-based logic with the following syntax: 
\begin{equation}
  \begin{aligned}
    \varphi ::  = & \top \mid \mu \mid \neg \varphi \mid \varphi_1 \wedge \varphi_2 \mid \\
                  & \F_\mathrm{I} \varphi \mid \G_\mathrm{I} \varphi \mid \varphi_1 \U_\mathrm{I} \varphi_2, 
  \end{aligned}
\end{equation}
where $\varphi$, $\varphi_1$, $\varphi_2$ are STL formulas and $\mathrm{I} := [a,b]$ with $0\leq a\leq b<\infty$ is a time interval. Here, $\neg$ and $\wedge$ are the logic operators “negation” and “conjunction”, respectively. $\F$, $\G$, and $\U$ are the temporal operators “eventually”, “always”, and “until”, respectively. 

The validity of an STL formula $\varphi$ with respect to a continuous-time signal $\boldsymbol{x}$ evaluated at time $t$ is defined inductively as follows: 
\begin{equation*}
  \begin{aligned}
      & (\boldsymbol{x},t)\models\mu && \Leftrightarrow g_\mu(\boldsymbol{x}(t))\geq0, \\
      & (\boldsymbol{x},t)\models\neg\varphi && \Leftrightarrow \neg((\boldsymbol{x},t)\vDash\varphi), \\
      & (\boldsymbol{x},t)\models\varphi_1\wedge\varphi_2 && \Leftrightarrow (\boldsymbol{x},t)\vDash\varphi_1\land(\boldsymbol{x},t)\vDash\varphi_2, \\
      & (\boldsymbol{x},t)\vDash\G_{[a,b]}\varphi && \Leftrightarrow \forall t\in[t+a,t+b],\text{s.t.}(\boldsymbol{x},t)\vDash\varphi, \\
      & (\boldsymbol{x},t)\vDash\F_{[a,b]}\varphi && \Leftrightarrow \exists t\in[t+a,t+b],\text{s.t.}(\boldsymbol{x},t)\vDash\varphi, \\
      & (\boldsymbol{x},t)\vDash\varphi_1\U_{[a,b]}\varphi_2 && \Leftrightarrow \exists t_1\in[t+a,t+b],\text{s.t.}(\boldsymbol{x},t_1)\vDash\varphi_2 \\
      & && \quad \ \wedge\forall t_2\in[t,t_1],(\boldsymbol{x},t_2)\vDash\varphi_1.
  \end{aligned}
\end{equation*}

An STL formula $\varphi$ is called a \textit{nested STL formula} if it contains more than one temporal operator; otherwise, it is called a \textit{non-nested STL formula} \citep[Definition 7]{yu2024continuous}. In this paper, we focus on two specific types of the nested STL formulas: $\F_{\mathrm{I}_1}\mathrm{G}_{\mathrm{I}_2}\varphi$ and $\mathrm{G}_{\mathrm{I}_1}\mathrm{F}_{\mathrm{I}_2}\varphi$, where $\I_1 := [a_1,b_1]$ and $\I_2 := [a_2,b_2]$. Specifically, $\F_{\mathrm{I}_1}\mathrm{G}_{\mathrm{I}_2}\varphi$ defines a time-constrained reach-stay task, which aims to reach the region of interest at a time instant $t \in \I_1$, and to stay within the region for the entire time interval $t \oplus \I_2$. $\mathrm{G}_{\mathrm{I}_1}\mathrm{F}_{\mathrm{I}_2}\varphi$ defines a time-constrained patrol task, which requires the system to enter the region of interest at least once every $b_2 - a_2$ period during the time interval $\I_1 \oplus \I_2$. The validity of these two nested STL formulas is defined as follows: 
\begin{equation*}
  \begin{aligned}
    (\boldsymbol{x},t)\vDash\mathrm{F}_{[a_1,b_1]}\mathrm{G}_{[a_2,b_2]}\varphi\Leftrightarrow & \exists t_1\in[t+a_1,t+b_1],\text{s.t.}\forall t_2\in \\
    & [t_1+a_2,t_1+b_2],(\boldsymbol{x},t_2)\vDash\varphi, \\
    (\boldsymbol{x},t)\vDash\mathrm{G}_{[a_1,b_1]}\mathrm{F}_{[a_2,b_2]}\varphi\Leftrightarrow & \forall t_1\in[t+a_1,t+b_1],\text{s.t.}\exists t_2\in \\
    & [t_1+a_2,t_1+b_2],(\boldsymbol{x},t_2)\vDash\varphi.
  \end{aligned}
\end{equation*}

\subsection{Problem Formulation}
\label{subsec-ProFor}
Consider the discrete-time mobile robot system with a sampling period of $\tau > 0$, described by 
\begin{equation}
  \label{equ-3}
  \boldsymbol{x}_{k+1}=f(\boldsymbol{x}_k, \boldsymbol{u}_k),\quad k \in \mathbb{N}_{\ge 0}, 
\end{equation}
where $\boldsymbol{x}_k \in \mathbb{X} \subset \mathbb{R}^n$ is the state including the position state $\boldsymbol{p}_k \in \mathbb{P} \subset \mathbb{R}^{\tilde{n}}$ and the non-position state $\boldsymbol{\theta}_k \in \Theta \subset \mathbb{R}^{n-\tilde{n}}$. Hence, $\mathbb{X} = \mathbb{P} \times \Theta$. $\boldsymbol{u}_k \in \mathbb{U} \subset \mathbb{R}^m$ is the control input. The function $f : \mathbb{R}^n \times \mathbb{R}^m \rightarrow  \mathbb{R}^n$ is assumed to be locally Lipschitz. For the mobile robot, a state sequence $\mathfrak{c}: \mathbb{N}_{\ge 0}  \rightarrow \mathbb{X}$ is called a trajectory if there exists a control sequence $\boldsymbol{u}: \mathbb{N}_{\ge 0}  \rightarrow \mathbb{U}$ such that \eqref{equ-3} holds for all $k \in \mathbb{N}_{\ge 0} $. Given a trajectory $\mathfrak{c}: \mathbb{N}_{\ge 0}  \rightarrow \mathbb{X}$, the corresponding position trajectory is defined as $\mathfrak{p} : \mathbb{N}_{\ge 0}  \rightarrow \mathbb{P} $, that is, $\mathfrak{p} := \mathtt{proj}(\mathfrak{c}, \mathbb{P})$. 

In the position space $\mathbb{P} \subset \mathbb{R}^{\tilde{n}}$, the set of all the obstacles is denoted by $\mathbb{O} := \cup_{\mathsf{l} \in \mathsf{L}} \mathcal{O}_{\mathsf{l}} \subset \mathbb{P}$ with a finite index set $\mathsf{L} \subset \mathbb{N}_{> 0} $. There exists a set $\Pi \subseteq \mathcal{AP}$ of propositions for \eqref{equ-3}. Each $\mu_{r} \in \Pi$ is associated with a set $\pi_{r} \subseteq \mathbb{P}$ such that $\mu_{r} = \top$ if $\mathtt{proj}(\mathfrak{c}, \mathbb{P}) \in \pi_{r}$, where $r \in \{1, \dots, \left | \Pi  \right | \}$ and $\left | \Pi  \right |$ is finite. In order to show the relation between $\mu_{r} \in \Pi$ and $\pi_{r} \subseteq \mathbb{P}$, we denote $\mathcal{R}(\mu_{r}):=\pi_{r}$. All these subsets are called the regions of interest, and are denoted by $\mathbb{D} := \cup_{r = 1}^{\left | \Pi \right | } \pi_{r}$. The mobile robot aims to accomplish the following STL fragment: 
\begin{subequations}
  \label{equ-4}
  \begin{align}
    \psi_i :: &= \mathrm{F}_{\I_i}\varphi_i \mid\mathrm{G}_{\I_i}\varphi_i,  \label{equ-4a}\\
    \psi_j :: &= \mathrm{F}_{\I_{j1}}\mathrm{G}_{\I_{j2}}\varphi_j \mid \mathrm{G}_{\I_{j1}}\mathrm{F}_{\I_{j2}}\varphi_j, \label{equ-4b} \\
    \Psi : &= (\wedge_{i=1}^{N}\psi_i) \wedge (\wedge_{j=N+1}^{N+M}\psi_j), \label{equ-4c}
  \end{align}
\end{subequations}
where $\psi_{i}$ is a non-nested sub-task, and $\psi_{j}$ is a nested sub-task. $N \in \mathbb{N}_{> 0} $ and $M \in \mathbb{N}_{> 0}$ are the numbers of the non-nested and nested sub-tasks, respectively. Note that $\varphi$ in \eqref{equ-4a} and \eqref{equ-4b} does not contain any temporal operators, i.e., $\varphi := \mu \mid \neg \mu$. If $\varphi = \mu$, then $\mathcal{R}(\varphi) := \mathcal{R}(\mu)$; otherwise, $\mathcal{R}(\varphi) := \mathbb{P} \setminus \mathcal{R}(\mu)$. $\psi_i$ in \eqref{equ-4a} can express the temporal operator $\mathrm{U}$, because $\varphi_{1} \mathrm{U}_{\I} \varphi_{2}$ can be written as $\mathrm{G}_{\I} \varphi_{1} \wedge \mathrm{F}_{\I} \varphi_{2}$; see \citep{yu2024continuous}. 

\begin{figure*}[htbp]
    \centering
    \subfigure[STL task decomposition]{
        \includegraphics[width=0.23\textwidth]{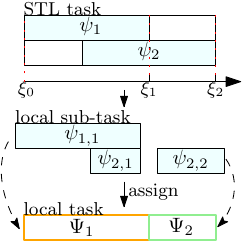}
        \label{fig-1a}
    }
    \hfill
    \subfigure[Local-to-global planning]{
        \includegraphics[width=0.3\textwidth]{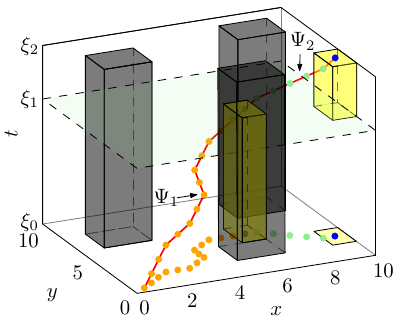}
        \label{fig-1b}
    }
    \hfill
    \subfigure[Trajectory optimization control]{
        \includegraphics[width=0.3\textwidth]{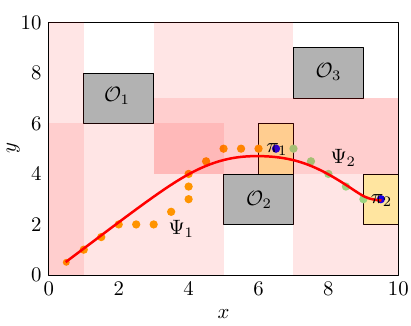}
        \label{fig-1c}
    }
    \caption{Overall architecture of the proposed approach. (a) The STL task $\Psi$ is decomposed into $\Psi_{1}$ and $\Psi_{2}$ with $\mathtt{uati}(\Psi_1) = [\xi_0, \xi_1]$ and $\mathtt{uati}(\Psi_2) = [\xi_1, \xi_2]$, respectively. (b) The sequences of local waypoints (orange and green points) satisfying $\Psi_{1}$ and $\Psi_{2}$ are generated and then synthesized into the sequence of global waypoints. Blue points are the satisfaction pairs with respect to $\Psi$. (c) Based on the sequence of global waypoints, a safe corridor (red blocks) is constructed. An optimal position trajectory (red line) satisfying $\Psi$ is derived.}
    \label{fig-1}
\end{figure*}

Similar to \citep{tan2024decomposition}, the duration of a sub-task $\psi_{\ell}$, $\ell \in \{1, \dots, N+M\}$, is termed as an application time interval (ATI), which is denoted by $\mathtt{ati}(\psi_{\ell})$. For $\psi_i$ in $\eqref{equ-4a}$, $\mathtt{ati}(\psi_i) = \mathrm{I}_i$. For $\psi_j$ in $\eqref{equ-4b}$, $\mathtt{ati}(\psi_j) = \mathrm{I}_{j1} \oplus \mathrm{I}_{j2}$. The duration of the STL task $\Psi$ in (4c) is called a union application time interval (UATI), which is denoted by $\mathtt{uati}(\Psi)$. For $\Psi$, $\mathtt{uati}(\Psi) := [0, th(\Psi)]$, where the time horizon is defined as $th(\Psi) := \max(\{\overline{\mathtt{ati}}(\psi_\ell)\}_{\ell = 1}^{N+M})$ from \citep[Definition 1]{sewlia2023cooperative}. 

\begin{assum}
  \label{assum-1}
  For each $\ell \in \{1, \ldots, N+M\}$, both $\underline{\mathtt{ati}}(\psi_{\ell})$ and $\overline{\mathtt{ati}}(\psi_{\ell})$ are the multiples of the sampling period $\tau > 0$.  
\end{assum}

\begin{assum}
  \label{assum-2}
  For each $\ell_1, \ell_2 \in \{N+1, \ldots, N+M\}$, $\mathtt{ati}^{\circ }(\psi_{\ell_1}) \cap \mathtt{ati}^{\circ }(\psi_{\ell_2}) = \emptyset$, for all $\ell_1 \ne \ell_2$.  
\end{assum}

Since the system (3) is discrete-time, we introduce Assumption \ref{assum-1} to facilitate the planning and control afterwards. Under this assumption, $th(\Psi)$ is a multiple of $\tau$, and the ceiling and floor functions (i.e., $\lceil\cdot\rceil$, $\lfloor\cdot\rfloor$) can be removed. For instance, $\F_{\I^{\prime}} \varphi$ with $\I^{\prime} := [a^{\prime}, b^{\prime}]$ can be redefined to correspond to the discrete-time setup. That is, $\I^{\prime}$ is redefined as $\I := [a, b]$, where $a := \tau \left \lceil a^{\prime} / \tau \right \rceil $ and $b := \tau \left \lfloor b^{\prime} / \tau \right \rfloor $. Moreover, the conservatism of Assumption 1 can be reduced by choosing a small sampling period.

Under Assumption \ref{assum-1}, this paper considers the system \eqref{equ-3} which is initially placed at a certain position $\boldsymbol{p}_0 \in \mathbb{P} \setminus \mathbb{O}$. Our goal is to propose a planning and control approach such that the STL task $\Psi$ in \eqref{equ-4c} is satisfied while guaranteeing the collision avoidance. The goal is transformed into the following optimization problem: 
\begin{subequations}
  \label{equ-5}
  \begin{align}
   \min_{\boldsymbol{x}_k, \boldsymbol{u}_k} \ & (\sum_{k=0}^{K-2} \left \| \boldsymbol{u}_{k+1} - \boldsymbol{u}_{k} \right \|_{\mathrm{Q}} + \sum_{k=0}^{K-1} \left \| \boldsymbol{x}_{k+1} - \boldsymbol{x}_{k} \right \|_{\mathrm{R}} ), \label{equ-5a} \\
   \text{s.t.} \ & \eqref{equ-3}, \label{equ-5b} \\
                 & \boldsymbol{x}_k \in \mathbb{X}, \quad \forall k\in\{0, \dots, K\}, \label{equ-5c} \\
                 &\boldsymbol{u}_k \in \mathbb{U}, \quad \forall k\in\{0, \dots, K-1\}, \label{equ-5d} \\
                 & \mathtt{proj}(\boldsymbol{x}_0, \mathbb{P}) = \boldsymbol{p}_0, \label{equ-5e} \\
                 & \mathtt{proj}(\mathfrak{c}, \mathbb{P}) \cap \mathbb{O} = \emptyset,  \label{equ-5f} \\
                 & \mathtt{proj}(\mathfrak{c}, \mathbb{P}) \models \Psi,  \label{equ-5g}                                         
  \end{align}
\end{subequations}
where $\mathrm{Q} \in \mathbb{R}^{m \times m}$ and $\mathrm{R} \in \mathbb{R}^{n \times n}$ are respectively two positive definite weight matrices, $\boldsymbol{x}_0$ is the initial state, and $K := th(\Psi) / \tau $. The first and second terms of the objective function in \eqref{equ-5a} correspond to maximizing the smoothness of the trajectory and minimizing the trajectory length, respectively. \eqref{equ-5c} and \eqref{equ-5d} are the state and input constraints, respectively. \eqref{equ-5f} and \eqref{equ-5g} enforce collision avoidance and satisfaction of the STL task, respectively.

\section{Methodology}
\label{sec-Methodology}

In order to solve the problem \eqref{equ-5}, a three-step planning and control approach is proposed in this section and is illustrated in Fig. \ref{fig-1}. The first step is to decompose an STL task into finite local tasks in Section \ref{subsec-DecomTask}, the second step is to derive the sequence of global waypoints and the corresponding satisfaction pair set in Section \ref{subsec-LocalToGlobal}, and the final step is to reformulate the problem \eqref{equ-5} to derive the optimal control strategy in Section \ref{subsec-OptimTraj}.

\subsection{STL Task Decomposition}
\label{subsec-DecomTask}
\begin{definition}[Local tasks]
  Consider an STL task $\Psi$ as in (4c). If there exist local tasks $\Psi_1, \dots, \Psi_{\mathcal{N}}$ such that for each $\imath_1, \imath_2 \in \{1, \dots, \mathcal{N} \}$,
  \begin{equation*}
    \left\{
      \begin{aligned}
      &\mathtt{uati}(\Psi_{\imath_1}) \subset  \mathtt{uati}(\Psi), \\
      &\mathtt{uati}^{\circ }(\Psi_{\imath_1}) \cap \mathtt{uati}^{\circ }(\Psi_{\imath_2}) = \emptyset, \imath_1 \ne \imath_2,\\
      &\mathtt{uati}(\Psi) = \cup_{\imath_1 = 1}^{\mathcal{N}} \mathtt{uati}(\Psi_{\imath_1}),
      \end{aligned}
    \right.
  \end{equation*}
  then both $\Psi_{\imath_1}$ and $\Psi_{\imath_2}$ are referred to as a local task of $\Psi$. The set of all local tasks of $\Psi$ is denoted as $\{\Psi_{1}, \dots, \Psi_{\mathcal{N}}\}$.
\end{definition}

In order to decompose the STL task \eqref{equ-4c} into multiple local tasks along the timeline, the first step is to determine the UATI of each local task, and the second step is to combine the local sub-tasks within each local UATI.

Based on the ATIs of all sub-tasks $\psi_{\ell}$, $\ell \in \{1, \dots, N+M\}$, the time horizons of all local tasks are defined as
\begin{equation}
  \label{equ-6}
  \mathbf{S}_{\xi} = \{\xi : \xi \notin \mathtt{ati}^{\circ} (\psi_{\ell}), \forall \ell \in \{N+1, \dots, N+M\} \}, 
\end{equation}
where $\xi \in \{\overline{\mathtt{ati}}(\psi_{\ell}) \}_{\ell=1}^{N+M}$. To represent all local UATIs, the set $(\{ \underline{\mathtt{uati}}(\Psi)\} \cup \mathbf{S}_{\xi})$ is arranged in an ascending order
\begin{equation}
  \label{equ-7}
  \mathbf{S}_{\xi}^{\text{ord}} := \mathtt{ord}(\{0\} \cup \mathbf{S}_{\xi}) = (\xi_{\imath})_{\imath = 0}^{\mathcal{N}}, \quad \xi_0 < \xi_1 < \dots < \xi_{\mathcal{N}},
\end{equation}
where $\mathcal{N} := \left | \mathbf{S}_{\xi} \right | \in \mathbb{N}_{> 0} $, $\xi_0 := 0$ and $\xi_{\mathcal{N}} := \overline{\mathtt{uati}}(\Psi)$. Therefore, there exist $\mathcal{N}$ local UATIs, and the $\imath$-th local UATI is denoted by $[\xi_{\imath-1}, \xi_{\imath}]$.

From (6) and (7), we partition the UATI of $\Psi$ into $\mathcal{N}$ local UATIs. That is,
\begin{equation}
  \label{equ-8}
  \mathtt{uati}(\Psi) = \cup_{\imath = 1}^{\mathcal{N}} [\xi_{\imath-1}, \xi_{\imath}].  
\end{equation}

\begin{remark}
  A sampling-based method can be used to generate a trajectory that satisfies a sub-task from either (4a) or (4b) within a time interval \citep{sewlia2023cooperative}. However, the method struggles to accomplish an STL task in (4c). Since a non-nested sub-task can be well decomposed into finite local ones \citep{tan2024decomposition, ren2022reachability}, we define each local UATI as in (8) such that $\left. 1 \right)$ within each local UATI, there exists a satisfied sub-task of the form (4a) or (4b); $\left. 2 \right)$ all nested sub-tasks in (4b) are not decomposed. \hfill $\square$
\end{remark}

Next, each sub-task $\psi_\ell$, $\ell \in \{1, \dots, N+M\}$, is assigned to the UATI that contains $\mathtt{ati}(\psi_{\ell})$. However, $\mathtt{ati}(\psi_{\ell})$ may span multiple local UATIs. Thus, $\psi_{\ell}$ is first decomposed into multiple local sub-tasks, and then these local sub-tasks are assigned to the corresponding UATIs. For decomposition, each $\psi_{\ell}$ has the following set:
\begin{equation}
  \mathbf{P}_{\ell} := \mathbf{S}_{\xi}^{\text{ord}} \cap \mathtt{ati}^{\circ}(\psi_{\ell}) \cup \{\underline{\mathtt{ati}}(\psi_{\ell}), \overline{\mathtt{ati}}(\psi_{\ell}) \}. 
\end{equation}
Its ordered sequence is denoted as 
\begin{equation}
  \mathtt{ord}(\mathbf{P}_{\ell}) := (\eta_{\ell, \jmath})_{\jmath = 0}^{\mathcal{M}_{\ell}}, \quad \eta_{\ell, 0} < \dots < \eta_{\ell, \mathcal{M}_{\ell}}, 
\end{equation}
where $\mathcal{M}_{\ell} := \left | \mathbf{P}_{\ell} \right | - 1 \in \mathbb{N}_{> 0} $, $\eta_{\ell, 0} := \underline{\mathtt{ati}}(\psi_{\ell})$ and $\eta_{\ell, \mathcal{M}_{\ell}} := \overline{\mathtt{ati}}(\psi_{\ell})$. Hence, from (9) and (10), $\mathtt{ati}(\psi_{\ell})$ is divided into $\mathcal{M}_{\ell}$ local ATIs. That is, 
\begin{equation}
  \mathtt{ati}(\psi_{\ell}) = \cup_{\jmath=1}^{\mathcal{M}_{\ell}} \left [ \eta_{\ell, \jmath-1}, \eta_{\ell, \jmath} \right ].
\end{equation}
Note that, from (8) and (11), for all $\jmath \in \{1, \dots, \mathcal{M}_{\ell}\}$, there exists a unique $\imath \in \{1, \dots, \mathcal{N}\}$ such that $\left [ \eta_{\ell, \jmath-1}, \eta_{\ell, \jmath} \right ] \subseteq \left [ \xi_{\imath-1}, \xi_{\imath} \right ]$. For assignment, each local UATI $[\xi_{\imath-1}, \xi_{\imath}]$ is associated with a set $\mathbf{S}_{\imath}$ to store the local sub-tasks. The following describes how a sub-task is decomposed into local ones, and how these local sub-tasks are assigned to the corresponding UATIs.

For a non-nested sub-task $\psi_{\ell}, \ell \in \{1, \dots, N\}$, the following three cases are addressed.
\begin{itemize}
  \item $\mathcal{M}_{\ell} = 1$ and $\psi_{\ell} = \G_{\I_{\ell}} \varphi_{\ell} \mid \F_{\I_{\ell}} \varphi_{\ell}$. In this case, $\mathtt{ati}(\psi_{\ell})$ does not span multiple local UATIs. Hence, $\psi_{\ell}$ is added to $\mathbf{S}_{\imath}$ if there exists $\imath \in \{1, \dots, \mathcal{N}\}$ such that $\mathtt{ati}(\psi_{\ell}) \subseteq \mathtt{uati}(\Psi_{\imath})$.
  \item $\mathcal{M}_{\ell} > 1$ and $\psi_{\ell} = \G_{\I_{\ell}} \varphi_{\ell}$. In this case, $\psi_{\ell}$ is decomposed as $\wedge_{\jmath = 1}^{\mathcal{M}_{\ell}} \psi_{\ell, \jmath}$ with 
        \begin{equation}
          \psi_{\ell, \jmath} = \G_{\I_{\ell, \jmath}} \varphi_{\ell}, \quad \I_{{\ell}, \jmath} = \left [ \eta_{\ell, \jmath-1}, \eta_{\ell, \jmath} \right ].  
        \end{equation}
        All local sub-tasks $\psi_{\ell, \jmath}$, $\jmath \in \{1, \dots, \mathcal{M}_{\ell} \}$, need to be satisfied since the temporal operator of $\psi_\ell$ is $\G$. Hence, each $\psi_{\ell, \jmath}$ is added to $\mathbf{S}_{\imath}$ if there exists $\imath \in \{1, \dots, \mathcal{N}\}$ such that $\mathtt{ati}(\psi_{\ell, \jmath}) \subseteq \mathtt{uati}(\Psi_{\imath})$.
  \item $\mathcal{M}_{\ell} > 1$ and $\psi_{\ell} = \F_{\I_{\ell}} \varphi_{\ell}$. In this case, $\psi_{\ell}$ is decomposed as $\vee_{\jmath = 1}^{\mathcal{M}_{\ell}} \psi_{\ell, \jmath}$ with
        \begin{equation}
          \psi_{\ell, \jmath} = \F_{\I_{\ell, \jmath}} \varphi_{\ell}, \quad \I_{\ell, \jmath} = \left [ \eta_{\ell, \jmath-1}, \eta_{\ell, \jmath} \right ].  
        \end{equation}
        Only one local sub-task $\psi_{\ell, \jmath}$, $\jmath \in \{1, \dots, \mathcal{M}_{\ell} \}$, needs to be satisfied since the temporal operator of $\psi_\ell$ is $\F$. However, which local sub-task $\psi_{\ell, \jmath}$ is satisfied is currently unknown and will be determined in the following subsection. Here, for a non-nested sub-task $\psi_{\ell}$ in the third case, we collect all its local sub-tasks $\psi_{\ell,\jmath}$ from (13) into a set. The set corresponding to the $\mathsf{n}$-th such sub-task is denoted by
        \begin{equation}
          \label{equ-14}
          \mathbf{D}_{\mathsf{n}} := \{\psi_{\ell, \jmath}^{\mathsf{n}} \}_{\jmath=1}^{\mathcal{M}_{\ell}}.
        \end{equation}
        Let $\bar{\mathsf{n}}$ be the total number of such sets $\mathbf{D}_{\mathsf{n}}$, and define the collection of these sets as: 
        \begin{equation}
          \label{equ-15}
          \mathbf{D} := \{\mathbf{D}_1, \dots, \mathbf{D}_{\bar{\mathsf{n}}}\}.
        \end{equation}
\end{itemize}

For a nested sub-task $\psi_{\ell}, \ell \in \{N+1, \dots, N+M\}$, from Remark 1, $\mathtt{ati}(\psi_{\ell})$ does not span multiple local UATIs. Hence, $\psi_{\ell}$ is assigned to $\mathbf{S}_{\imath}$ if there exists $\imath \in \{1, \dots, \mathcal{N}\}$ such that $\mathtt{ati}(\psi_{\ell}) \subseteq \mathtt{uati}(\Psi_{\imath})$. 

After all sub-tasks in (4a) and (4b) are decomposed as above, let $\bar{\mathsf{m}}_{\imath} := \left | \mathbf{S}_{\imath}  \right | $, and each local sub-task in $\mathbf{S}_{\imath}$ is denoted by $\psi_{\imath, \mathsf{m}}$. For each $\mathsf{m} \in \{1, \dots, \bar{\mathsf{m}}_{\imath}\}$, the specific form is given below. 
\begin{equation}
  \label{equ-16}
  \begin{aligned}
    \psi_{\imath, \mathsf{m}} : =& \F_{\I_{\imath, \mathsf{m}}} \varphi_{\imath, \mathsf{m}} \mid \F_{\I_{\imath, \mathsf{m}1}} \G_{\I_{\imath, \mathsf{m}2}} \varphi_{\imath, \mathsf{m}} \mid \\
                      & \G_{\I_{\imath, \mathsf{m}}} \varphi_{\imath, \mathsf{m}} \mid \G_{\I_{\imath, \mathsf{m}1}} \F_{\I_{\imath, \mathsf{m}2}} \varphi_{\imath, \mathsf{m}}.
  \end{aligned}
\end{equation}
Each local task $\Psi_{\imath}$, $\imath \in \{1, \dots, \mathcal{N}\}$, is constructed by taking the conjunction of all $\psi_{\imath, \mathsf{m}} \in \mathbf{S}_{\imath}$, and is expressed as 
\begin{equation}
  \label{equ-17}
  \Psi_{\imath} = \wedge_{\psi_{\imath, \mathsf{m}} \in \mathbf{S}_{\imath}} \psi_{\imath, \mathsf{m}}, \quad \mathtt{uati}(\Psi_{\imath}) = \left [ \xi_{\imath-1}, \xi_{\imath} \right ]. 
\end{equation}
Finally, $\Psi$ is decomposed into $\mathcal{N}$ local tasks. That is,  
\begin{equation}
  \Psi = \wedge_{\imath=1}^{\mathcal{N}} \Psi_{\imath}, \quad \mathtt{uati}(\Psi) = \cup_{\imath = 1}^{\mathcal{N} } \mathtt{uati}(\Psi_{\imath}).
\end{equation}
Therefore, we have the following lemma.

\begin{lem}
  \label{lem-1}
  Given an STL task $\Psi$ in (4c) that can be decomposed as $\wedge_{\imath=1}^{\mathcal{N}} \Psi_{\imath}$ with $\mathtt{uati}(\Psi) = \cup_{\imath = 1}^{\mathcal{N} } \mathtt{uati}(\Psi_{\imath})$, $\Psi$ is satisfied if all $\Psi_{\imath}$ are satisfied.
\end{lem}

\begin{remark}
  The above decomposition yields a reformulation only for sub-tasks of the form $\psi_{\ell}=\F_{\I_{\ell}}\varphi_{\ell}$ with $\mathcal{M}_{\ell}>1$, because the decomposition reduces their ATIs. For all other sub-tasks, the decomposition preserves equivalence with the original sub-tasks. \hfill $\square$
\end{remark}

\begin{algorithm2e}[t]
\label{alg-1}
\caption{STL Satisfaction ($\mathtt{STLSat}$)}
\KwIn{ $\mathfrak{q}, \psi$ }
\KwOut{ $(\mathcal{B}_{\imath, \mathsf{m}}, \Upsilon_{\imath, \mathsf{m}})$ }

$\mathcal{B}_{\imath, \mathsf{m}} = \bot$ \;
$\Upsilon_{\imath, \mathsf{m}} = \{ \}$ \;

\If{ $\psi = \F_{\I} \varphi$ }{
  $\mathsf{T}_1 = \mathbf{T} \cap \I$ \;
  \For{ $i = 1 : \left | \mathsf{T}_1 \right |$ }{
    \If{$\mathfrak{q}(\mathsf{T}_1(i)) \in \mathcal{R}(\varphi) $}{
      $\mathcal{B}_{\imath, \mathsf{m}} = \top$ \;
      $\Upsilon_{\imath, \mathsf{m}} = \left (\mathsf{T}_1(i), \mathcal{R}(\varphi) \right )$ \;
    }
  }
}

\If{$\psi = \G_{\I} \varphi $}{
  $\mathsf{T}_1 = \mathbf{T} \cap \I$ \;
  \For{$i = 1 : \left | \mathsf{T}_1 \right |$}{
    \If{$\mathfrak{q}(\mathsf{T}_1(i)) \notin \mathcal{R}(\varphi)$}{
      \textbf{break} \;
    }
  }
  $\mathcal{B}_{\imath, \mathsf{m}} = \top$ \;
  \For{$i = 1 : \left | \mathsf{T}_1 \right |$}{
    $\Upsilon_{\imath, \mathsf{m}} = \Upsilon_{\imath, \mathsf{m}} \cup \left ( \mathsf{T}_1(i), \mathcal{R}(\varphi) \right )$ \;
  }
}

\If{$\psi = \F_{\I_1} \G_{\I_2} \varphi $}{
  $\mathsf{T}_1 = \mathbf{T} \cap \I_1$ \;
  \For{$i = 1 : \left | \mathsf{T}_1 \right |$}{
    $\mathsf{T}_2 = \mathbf{T} \cap (\mathsf{T}_1(i) \oplus \I_2 )$ \;
    \If{ $\mathfrak{q}(\mathsf{T}_2) \subseteq \mathcal{R}(\varphi)$ }{
      $\mathcal{B}_{\imath, \mathsf{m}} = \top$ \;
      \For{$j = 1 : \left | \mathsf{T}_2 \right | $}{
        $\Upsilon_{\imath, \mathsf{m}} = \Upsilon_{\imath, \mathsf{m}} \cup \left ( \mathsf{T}_2(j), \mathcal{R}(\varphi) \right )$ \;
      }
    }
  }
}

\If{$\psi = \G_{\I_1} \F_{\I_2} \varphi $}{
  $\mathsf{T}_2 = \{ \}$ \;
  $\mathsf{T}_1 = \mathbf{T} \cap (\I_1 \oplus \I_2)$ \;
  \For{$i = 1 : \left | \mathsf{T}_1 \right |$}{
    \If{$\mathfrak{q}(\mathsf{T}_1(i)) \in \mathcal{R}(\varphi)$}{
      $\mathsf{T}_2 = \mathsf{T}_2 \cup \mathsf{T}_1(i)$ \;
    }
  }
  \For{$j = 1 : \left | \mathsf{T}_2 \right | - 1$}{
    \If{$\mathsf{T}_2(j+1) - \mathsf{T}_2(j) > e$}{
      \textbf{break} \;
    }
  }
  \If{$\mathsf{T}_2(1) - \underline{\mathtt{ati}}(\psi) > e$ \textnormal{ or } $\overline{\mathtt{ati}}(\psi) - \mathsf{T}_2(\left | \mathsf{T}_2 \right |) > e $}{
    \textbf{break} \;
  }
  $\mathcal{B}_{\imath, \mathsf{m}} = \top$ \;
  \For{$j = 1 : \left | \mathsf{T}_2 \right |$}{
    $\Upsilon_{\imath, \mathsf{m}} = \Upsilon_{\imath, \mathsf{m}} \cup \left ( \mathsf{T}_2(j), \mathcal{R}(\varphi) \right )$ \;
  }

}

\end{algorithm2e}

\subsection{Local-to-Global Planning}
\label{subsec-LocalToGlobal}
From Lemma 3, the satisfaction of each local sub-task ensures the satisfaction of the STL task. To this end, we first introduce the STL satisfaction checking of the local sub-tasks in \eqref{equ-16}, then generate a sequence of local waypoints to satisfy the local task in \eqref{equ-17}. Finally, a local-to-global strategy is applied to synthesize a sequence of global waypoints that satisfies the STL task.

If there exist a time interval $[\mathfrak{t}_1,\mathfrak{t}_2]$ and a sampling of period $\tau > 0$ such that $\tfrac{\mathfrak{t}_2 - \mathfrak{t}_1 }{\tau} \in \mathbb{N}_{\ge 0}$, then $\mathbf{T} := \{ t_\mathsf{j} : t_\mathsf{j} = \mathfrak{t}_1 + \mathsf{j}\tau, \mathsf{j} = 0,1,\ldots,\mathsf{N} \}$, and a sequence of points can be defined as
\begin{equation}
  \label{equ-19}
  \mathfrak{q} := \{ (p_\mathsf{j}, t_\mathsf{j}) : t_\mathsf{j} \in \mathbf{T} \},
\end{equation}
where $\mathsf{N} := \tfrac{\mathfrak{t}_2 - \mathfrak{t}_1 }{\tau}$. The position of $\mathfrak{q}$ at time instant $t \in \mathbf{T}$ is expressed as $\mathfrak{q}(t) := p$. Let $\mathsf{T} \subset \mathbf{T} $, $\mathfrak{q}(\mathsf{T}) := \{\mathfrak{q}(t): t \in \mathsf{T} \}$. In order to verify whether $\mathfrak{q} \models \psi_{\imath, \mathsf{m}}$, we introduce the following definition and algorithm. 

\begin{definition}[Satisfaction pair set]
  Consider a sequence of points $\mathfrak{q}$ from \eqref{equ-19}, a local sub-task $\psi_{\imath, \mathsf{m}}$ in \eqref{equ-16}, and $\mathsf{T}_1 := \mathbf{T} \cap \mathtt{ati}(\psi_{\imath, \mathsf{m}})$. If $\mathfrak{q} \models \psi_{\imath, \mathsf{m}}$, then there exist some time instants $\mathsf{T}_2 \subseteq \mathsf{T}_1$ such that $\mathfrak{q}(\mathsf{T}_2)$ lie within $\mathcal{R}(\varphi_{\imath, \mathsf{m}})$. We refer to $\Upsilon := \{(t, \mathcal{R}(\varphi_{\imath, \mathsf{m}})): t \in \mathsf{T}_2 \}$ as the satisfaction pair set of $\mathfrak{q}$ with respect to $\psi_{\imath, \mathsf{m}}$, and each $(t, \mathcal{R}(\varphi_{\imath, \mathsf{m}})) \in \Upsilon$ is called a satisfaction pair.
\end{definition}

Algorithm 1 presents how to determine if $\mathfrak{q} \models \psi_{\imath, \mathsf{m}}$. It outputs a Boolean value $\mathcal{B}_{\imath, \mathsf{m}} \in \{\top, \bot \}$, and the corresponding satisfaction pair set $\Upsilon_{\imath, \mathsf{m}}$. Depending on the type of $\psi_{\imath, \mathsf{m}}$, four different cases are considered. For convenience, we temporarily omit the index of the local sub-task $\psi_{\imath, \mathsf{m}}$ in the following cases.
\begin{itemize}
  \item $\psi = \F_{\I} \varphi$. If there exists $t \in \mathsf{T}_1 := \mathbf{T} \cap \I$ such that $\mathfrak{q}(t) \in \mathcal{R}(\varphi)$, then $\mathfrak{q} \models \psi$. The satisfaction pair set is $\Upsilon_{\imath, \mathsf{m}} := \left(t, \mathcal{R}(\varphi) \right)$. 
  \item $\psi = \G_{\I} \varphi$. If there exists $\mathsf{T}_1 := \mathbf{T} \cap \I$ such that $\mathfrak{q}(\mathsf{T}_1) \subseteq \mathcal{R}(\varphi)$, then $\mathfrak{q} \models \psi$. The satisfaction pair set is $\Upsilon_{\imath, \mathsf{m}} := \{\left( t, \mathcal{R}(\varphi) \right): t \in \mathsf{T}_1 \}$. 
  \item $\psi = \F_{\I_1} \G_{\I_2} \varphi$. If there exists $t_1 \in \mathsf{T}_1 := \mathbf{T} \cap \I_1$ such that $\mathfrak{q}(\mathsf{T}_2) \subseteq \mathcal{R}(\varphi)$ with $\mathsf{T}_2 := \mathbf{T} \cap (t_1 \oplus \I_2)$, then $\mathfrak{q} \models \psi$. The satisfaction pair set is $\Upsilon_{\imath, \mathsf{m}} := \{\left( t, \mathcal{R}(\varphi) \right): t \in \mathsf{T}_2 \}$.
  \item $\psi = \G_{\I_1} \F_{\I_2} \varphi$ with $\I_1 := [a_1, b_1]$ and $\I_2 := [a_2, b_2]$. Let $e := b_2 - a_2$, $\mathsf{T}_1 := \mathbf{T} \cap (\I_1 \oplus \I_2) $, and $\mathsf{T}_2 := \{t \in \mathsf{T}_1: \mathfrak{q}(t)\in\mathcal{R}(\varphi)\}$. If for all $t_{i}, t_{i+1} \in \mathsf{T}_2$ such that $(t_{i+1} - t_{i}) \le e $, $\mathsf{T}_2 (1) - \underline{\mathtt{ati}}(\psi) \le e$, and $\overline{\mathtt{ati}}(\psi) - \mathsf{T}_2(\left | \mathsf{T}_2 \right | ) \le e $, then $\mathfrak{q} \models \psi$. The satisfaction pair set is $\Upsilon_{\imath, \mathsf{m}} := \{\left( t, \mathcal{R}(\varphi) \right): t \in \mathsf{T}_2\}$.
\end{itemize}

With the proposed Algorithm 1, we next introduce the following two key ingredients for Algorithm 2. 
\begin{itemize}
  \item The first ingredient is to introduce the algorithm used to generate the local path. The Goal-Biased Rapidly-exploring Random Tree (Goal-Biased RRT) is an incrementally expanding tree denoted by $\mathcal{T}= (\mathcal{V},\mathcal{E})$ \citep{urmson2003approaches, sewlia2023cooperative}, where $\mathcal{V}$ and $\mathcal{E}$ represent the sets of vertices and edges, respectively. In this work, each vertex in the set $\mathcal{V}$ is given by $q_i := (p_i, t_i) \in (\mathbb{P} \setminus \mathbb{O}) \times \mathbb{R}_{\ge 0}$. $\mathtt{pos}(q_i) := p_i$ and $\mathtt{time}(q_i) := t_i$. The tree $\mathcal{T}$ is iteratively constructed through the three primary functions: $\mathtt{Sample}$, $\mathtt{Nearest}$, and $\mathtt{Steer}$. The specific forms of these three functions will be presented later.
  \item The second ingredient is to address the assignment of the local sub-tasks in (14). Specifically, for each $\mathbf{D}_{\mathsf{n}} \in \mathbf{D}$, we examine all local sub-tasks $\psi_{\ell, \jmath}^{\mathsf{n}} \in \mathbf{D}_{\mathsf{n}}$, $\jmath \in \{1, \dots, \mathcal{M}_{\ell} \}$. If the following two conditions both hold, then $\psi_{\ell, \mathcal{M}_{\ell}}^{\mathsf{n}}$ is assigned to $\mathbf{S}_{\imath}$ to update $\Psi_{\imath}$. The first condition holds if all $\psi_{\ell,\jmath}^{\mathsf{n}}$, $\jmath \in \{1, \dots, \mathcal{M}_{\ell} -1 \}$, are not satisfied. Their satisfaction is checked using Algorithm 1 against the previously generated sequences of local waypoints. The second condition holds if $\mathtt{ati}(\psi_{\ell, \mathcal{M}_{\ell}}^{\mathsf{n}}) \subseteq \mathtt{uati}(\Psi_{\imath})$.
\end{itemize}

\begin{algorithm2e}[t]
\label{alg-2}
\caption{Local Planning}
\KwIn{ $\Psi_{\imath}, q_{\text{init}}$ }
\KwOut{ $\Gamma_{\imath}, \Upsilon_{\imath}$ }

\textnormal{LoopVal = True} \;
\While{ \textnormal{LoopVal} }{
  $\mathcal{B}_{\imath} = \top$ \;
  $\bar{\mathcal{P}}_\imath = \{ \}$ \;
  $q_\text{start} = q_{\text{init}}$ \;

  \While{$T_{\text{cur}} < \xi_{\imath}$}{
    $\text{Tar} = \mathtt{select}(\text{TarList})$ \;
    $\bar{\mathcal{P}}_\text{c} = \mathtt{GenTree}(q_\text{start}, \text{Tar})$ \;
    $\bar{\mathcal{P}}_\imath = \bar{\mathcal{P}}_\imath \cup \bar{\mathcal{P}}_\text{c}$ \;
    $q_\text{start} = [\bar{\mathcal{P}}_\text{c}]_\text{e}$ \;
    $T_{\text{cur}} = \mathtt{time}( [\bar{\mathcal{P}}_\text{c}]_\text{e})$ \;
  }

  $\mathcal{P}_\imath = \mathtt{Con}(\bar{\mathcal{P}}_\imath)$ \;
  
  \For{$\mathsf{m} = 1: \bar{\mathsf{m}}_{\imath} $}{
    $(\mathcal{B}_{\imath, \mathsf{m}}, \Upsilon_{\imath, \mathsf{m}}) = \mathrm{STLSat}(\mathcal{P}_{\imath}, \psi_{\imath, \mathsf{m}})$ \;
    
    \If{$\mathcal{B}_{\imath, \mathsf{m}} = \bot$}{
      $\mathcal{B_{\imath}} = \bot$ \;
    }

  }

  \If{$\mathcal{B}_{\imath} = \top$}{
    $\Gamma_{\imath} = \mathcal{P}_{\imath}$ \;
    $\Upsilon_{\imath} = \{\Upsilon_{\imath, \mathsf{m}} \}_{\mathsf{m} = 1}^{\bar{\mathsf{m}}_{\imath}}$ \;
    \textnormal{LoopVal = False} \;
  }

}

\end{algorithm2e}

Algorithm 2 outputs the sequence of local waypoints $\Gamma_{\imath}$ satisfying the local task $\Psi_{\imath}$, and the corresponding satisfaction pair set $\Upsilon_{\imath}$. It has the following three stages: $\left. 1 \right)$ generating a local path $\bar{\mathcal{P}}_{\imath}$ (lines 4-11), $\left. 2 \right)$ discretizing the path into a sequence of points $\mathcal{P}_{\imath}$ (line 12), and $\left. 3 \right)$ verifying whether $\mathcal{P}_{\imath}$ satisfies all local sub-tasks $\psi_{\imath, \mathsf{m}} \in \Psi_{\imath}$ (lines 13-19). If the verification succeeds, then $\Gamma_{\imath} := \mathcal{P}_{\imath}$, and $\Upsilon_{\imath}$ is determined; otherwise, the process iterates until the outputs are derived.

\begin{figure}[htbp]
    \centering
    \includegraphics[width=0.3\textwidth]{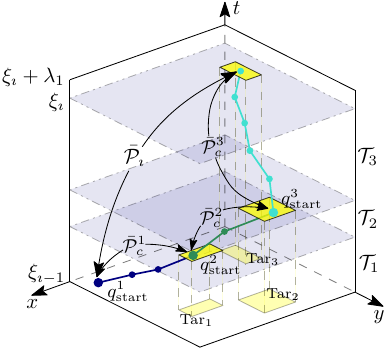}
    \caption{The generation of the local path. The local path $\bar{\mathcal{P}}_{\imath}$ consists of three current paths, denoted as $\bar{\mathcal{P}}_{c}$. Each $\bar{\mathcal{P}}_{c}$ is obtained by constructing a tree $\mathcal{T}_c$ rooted at the $q_{\text{start}}$ and expanding it toward the target region (yellow rectangle).}
    \label{fig-2}
\end{figure}

The first stage is as follows; see Fig. \ref{fig-2}. In line 4, $\bar{\mathcal{P}}_\imath$ is introduced to store the local path. In line 5, if $\imath = 1$, then $q_{\text{init}} := (\boldsymbol{p}_0, 0)$; otherwise, $q_{\text{init}}$ is the last waypoint of $\Gamma_{\imath-1}$. In line 7, the specific form of ${\text{TarList}}$ is defined as 
\begin{equation}
  \text{TarList} := \{\mathcal{R}(\varphi_{\imath, \mathsf{m}}): \forall \psi_{\imath, \mathsf{m}} \in \mathbf{S}_{\imath} \}. 
\end{equation}
The function $\mathtt{select}$ is as follows. The region of interest is sequentially selected from the ${\text{TarList}}$ as a target region $\text{Tar}$. If the temporal operator corresponding to the selected $\text{Tar}$ is $\F$, the region of interest will be removed from the ${\text{TarList}}$. In line 8, after determining the starting point and the target region, a current path $\bar{\mathcal{P}}_{\text{c}}$ is obtained through $\mathtt{GenTree}$ operation. Specifically, the $\mathtt{GenTree}$ operation first constructs a Goal-Biased RRT tree $\mathcal{T}_c := (\mathcal{V}_c, \mathcal{E}_c )$ by iteratively using the following three functions, and then returns a path once the tree is completed. 
\begin{itemize}
  \item The function $\mathtt{Sample}$ samples a random point $q_{\text{samp}} = (p_{\text{samp}}, t_{\text{samp}}) \in (\mathbb{P} \setminus \mathbb{O}) \times [\xi_{\imath-1}, \xi_{\imath} + \lambda_1 ]$, where $\lambda_1  \in \mathbb{R}_{\ge 0} $. The function $\mathtt{rand}$ that selects an element uniformly from a given set, which is used to derive 
    \begin{equation*}
      p_{\text{samp}} := 
      \begin{cases}
        \mathtt{rand}(\text{Tar}),&\text{with probability } \delta_{\text{tar}}, 
        \\
        \mathtt{rand}(\mathbb{P} \setminus \mathbb{O} ),&\text{with probability } 1-\delta_{\text{tar}}, 
      \end{cases}
    \end{equation*}
    and $t_{\text{samp}} := \mathtt{rand}([\xi_{\imath-1}, \xi_{\imath}+\lambda_1])$, where $\delta_{\text{tar}} \in \left [ 0,1 \right )$ represents the probability of such sampling within $\text{Tar}$. 
  \item The function $\mathtt{Nearest}$ identifies the nearest point $q_\text{near} = (p_\text{near}, t_\text{near}) \in \mathcal{V}_c $ as follows: 
        \begin{equation*}
          \begin{aligned}
          q_\text{near} : =  &\arg \min_{q_i \in \mathcal{V}_c } \left \| \mathtt{pos}(q_i) - p_\text{samp} \right \|,\\
                                & \text{s.t.} \quad \mathtt{time}(q_i) < t_\text{samp.} 
          \end{aligned}
        \end{equation*}
  \item The function $\mathtt{Steer}$ generates the new point $q_{\text{new}} = (p_{\text{new}}, t_{\text{new}})$ as follows: 
        \begin{equation*}
          p_\text{new} := 
          \begin{cases}
          p_\text{samp}, \text{ if } \left \| p_\text{samp} - p_\text{near} \right \| \le \Delta h,\\
          p_\text{near} + \tfrac{ \Delta h (p_\text{samp} - p_\text{near})}{\left \| p_\text{samp} - p_\text{near} \right \|}, \text{ otherwise },
          \end{cases}
        \end{equation*}
        where $t_\text{new} := \min \{t_\text{near} + \delta_t, t_\text{samp} \}$ with $\delta_t \in \mathbb{R}_{>0} $, and $\Delta h \in \mathbb{R}_{>0}$ is the step size of the Goal-Biased RRT.
\end{itemize}
In lines 10-11, the starting point $q_\text{start}$ and current time $T_\text{cur}$ are updated, where $[\bar{\mathcal{P}}_\text{c}]_\text{e}$ denotes the last point of $\bar{\mathcal{P}}_{\text{c}}$. When $T_\text{cur} \ge \xi_{\imath}$, a local path $\bar{\mathcal{P}}_{\imath} := \{(p_{\imath, \mathsf{i}}, t_{\imath, \mathsf{i}})\}_{\mathsf{i} = 0}^{\mathsf{M}_{\imath} }$ is derived, which is composed of multiple segments of the current path. 

The second stage is to ensure that all points are uniformly discretized in time. Specifically, the local path is discretized into a sequence of points via the function $\mathtt{Con}$. These points are denoted as $\mathcal{P}_{\imath} := \{(p_{\imath, \mathsf{j}}, t_{\imath, \mathsf{j}})\}_{\mathsf{j} = 0}^{\mathsf{N}_{\imath} }$ with $\mathsf{N}_{\imath} := (\xi_{\imath} - \xi_{\imath-1})/\tau$. Note that Assumption \ref{assum-1} is used to ensure $\mathsf{N}_{\imath} \in \mathbb{N}_{>0}$. For each $(p_{\imath, \mathsf{j}}, t_{\imath, \mathsf{j}}) \in \mathcal{P}_{\imath}$, $t_{\imath, \mathsf{j}} : = \xi_{\imath-1} + \mathsf{j} \tau$, and then $p_{\imath, \mathsf{j}} : = p_{\imath, \mathsf{i}} + \tfrac{(t_{\imath, \mathsf{j}} - t_{\imath, \mathsf{i}})}{(t_{\imath, \mathsf{i+1}} - t_{\imath, \mathsf{i}})}(p_{\imath, \mathsf{i+1}} - p_{\imath, \mathsf{i}})$ if $t_{\imath, \mathsf{j}} \in \left [ t_{\imath, \mathsf{i}}, t_{\imath, \mathsf{i+1}} \right )$.

The third stage verifies the satisfaction of $\Psi_{\imath}$ by $\mathcal{P}_{\imath}$. In lines 13-14, for each local sub-task $\psi_{\imath,\mathsf{m}} \in \mathbf{S}_{\imath}$, Algorithm 1 is applied to $\mathcal{P}_{\imath}$ to obtain the Boolean value $\mathcal{B}_{\imath, \mathsf{m}}$ and the satisfaction pair set $\Upsilon_{\imath, \mathsf{m}}$ with respect to $\psi_{\imath, \mathsf{m}}$. In lines 15-19, if all  $\mathcal{B}_{\imath, \mathsf{m}}$ are $\top$, then we derive the sequence of local waypoints $\Gamma_{\imath} := \mathcal{P}_{\imath}$, and the satisfaction pair set  $\Upsilon_{\imath} := \{\Upsilon_{\imath,\mathsf{m}}\}_{\mathsf{m}=1}^{\bar{\mathsf{m}}_{\imath}}$ with respect to $\Psi_{\imath}$.

For each local task $\Psi_{\imath}$, $\imath \in \{1, \dots, \mathcal{N}\}$, we use Algorithm 2 to derive the sequence of local waypoints $\Gamma_{\imath}$ that satisfies $\Psi_{\imath}$, and the satisfaction pair set $\Upsilon_{\imath}$ with respect to $\Psi_{\imath}$. From Lemma 3, we apply a local-to-global strategy to derive the sequence of global waypoints $\Gamma$ that satisfies the STL task $\Psi$. $\Gamma$ and the satisfaction pair set $\Upsilon$ with respect to $\Psi$ are defined as
\begin{equation}
  \left\{\begin{matrix}
  \begin{aligned}
  \Gamma    :&= \{(\boldsymbol{p}_0, 0)\} \cup (\cup_{\imath  = 1}^{\mathcal{N}}(\{(p_{\imath, \mathsf{j}}, t_{\imath, \mathsf{j}})\}_{\mathsf{j} = 1}^{\mathsf{N}_{\imath} })),\\
  \Upsilon  :&= \{\Upsilon_{\imath}\}_{\imath  = 1}^{\mathcal{N}},
  \end{aligned}
  \end{matrix}\right.
\end{equation}
where $ \left | \Gamma \right | = K+1 $. Note that the value of $K$ here is equal to that in (5a). The $k$-th global waypoint is denoted as $\Gamma(k)$. 

\subsection{Trajectory Optimization Control}
\label{subsec-OptimTraj}

In this subsection, the sequence of global waypoints $\Gamma$ is used to construct a safe corridor. The safe corridor and satisfaction pair set $\Upsilon$ are then used to replace the constraints (5f) and (5g) of the optimization problem (5). By solving the reformulated problem, we derive an optimal control sequence such that the position trajectory satisfies the STL task $\Psi$.

\begin{algorithm2e}[t]
\label{alg-3}
\caption{Construct Safe Corridor}
\KwIn{$\Gamma, \mathbb{P}$}
\KwOut{$\mathbf{SC}$}
$\mathcal{S}_{0} = \mathtt{SafeCor}(\Gamma(0), \mathbb{P})$ \;
$\mathbf{SC} = \{\mathcal{S}_{0} \}$ \;
\For{ $k = 1:K$ }{
  \If{ $\Gamma(k) \in \mathcal{S}_{k-1}$ }{
    $\mathcal{S}_k = \mathcal{S}_{k-1}$ \;
    $\mathbf{SC} = \mathbf{SC} \cup \mathcal{S}_k$ \;
  }
  \Else{
    $\mathcal{S}_k = \mathtt{SafeCor}(\Gamma(k), \mathbb{P})$ \;
    $\mathbf{SC} = \mathbf{SC} \cup \mathcal{S}_k$ \;
  }
}
\end{algorithm2e}

Algorithm 3 is designed to generate a sequence of safe corridors based on $\Gamma$ and $\mathbb{P}$. Specifically, each waypoint $\Gamma(k)$ corresponds to a safe corridor $\mathcal{S}_k$. If $\Gamma(k) \in \mathcal{S}_{k-1}$, then $\mathcal{S}_k := \mathcal{S}_{k-1}$; otherwise, the function $\mathtt{SafeCor}$ constructs the safe corridor $\mathcal{S}_k$. The complete formulation of $\mathtt{SafeCor}$ can be found in \citep[Algorithm 1]{park2020efficient} and \citep[Algorithm 2]{9293348}. The collection of all safe corridors is denoted by $\mathbf{SC} = \{\mathcal{S}_k : k = 0, \dots, K\}$.

Based on the safe corridor $\mathbf{SC}$ and the satisfaction pair set $\Upsilon$, we reformulate the problem \eqref{equ-5} as  
\begin{subequations}
  \begin{align}
    \min_{\boldsymbol{x}_k, \boldsymbol{u}_k} \ & (\sum_{k=0}^{K-2} \left \| \boldsymbol{u}_{k+1} - \boldsymbol{u}_{k} \right \|_{\mathrm{Q}} + \sum_{k=0}^{K-1} \left \| \boldsymbol{x}_{k+1} - \boldsymbol{x}_{k} \right \|_{\mathrm{R}} ), \label{equ-20a} \\
    \text{s.t.} \ & (5\text{b}) - (5\text{e}), \label{equ-20b} \\
                  & \mathtt{proj}(\boldsymbol{x}_k, \mathbb{P}) \in \mathcal{S}_k,\quad\forall k \in \{0, \dots, K \}, \label{equ-20c} \\
                  & \mathtt{proj}(\boldsymbol{x}_k, \mathbb{P}) \in \mathbf{r}_s, \quad \forall (k\tau, \mathbf{r}_s) \in \Upsilon, \label{equ-20d}                                     
  \end{align}
\end{subequations}
where $\mathbf{r}_s \in \mathbf{R} := \{\pi_r \in \mathbb{D}: (k \tau, \pi_r) \in \Upsilon \} $ with $s \in \{1, \dots, \left | \mathbf{R} \right |  \}$. The positions of the optimal trajectory are limited in the safe corridor $\mathbf{SC}$ in the constraint \eqref{equ-20c}. In constraint \eqref{equ-20d}, for all $k \in \{0, \dots, K\}$, if $(k\tau, \mathbf{r}_s) \in \Upsilon$, then the position of the $k$-th point is constrained to lie within the regions of interest $\mathbf{r}_s$. From Definition 2, the STL task $\Psi$ is accomplished as long as the trajectory satisfies the constraint (22d).

\begin{figure*}[htbp]
    \centering
    \subfigure[]{
        \includegraphics[width=0.3\textwidth]{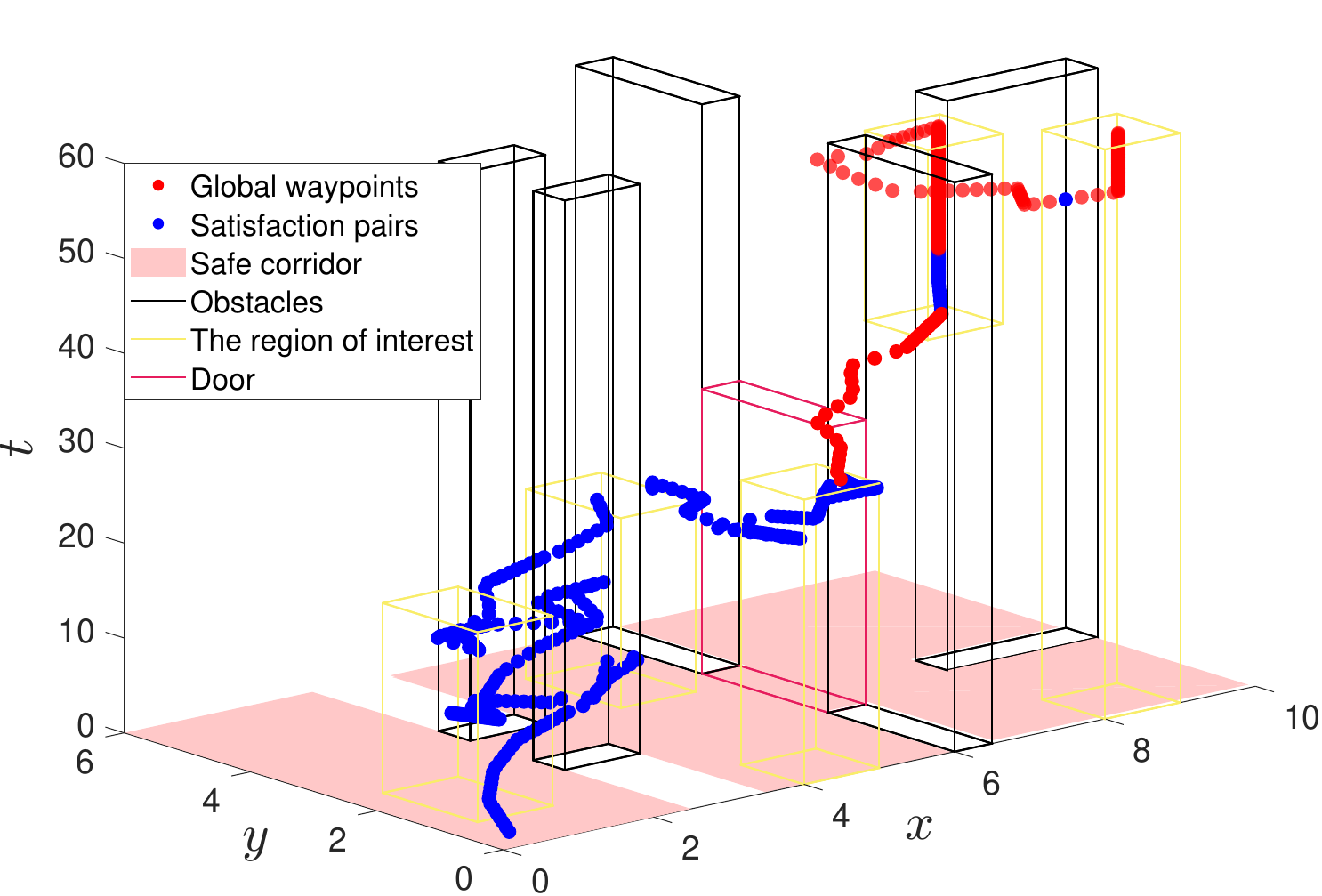}
        \label{fig-3a}
    }
    \hfill
    \subfigure[]{
        \includegraphics[width=0.334\textwidth]{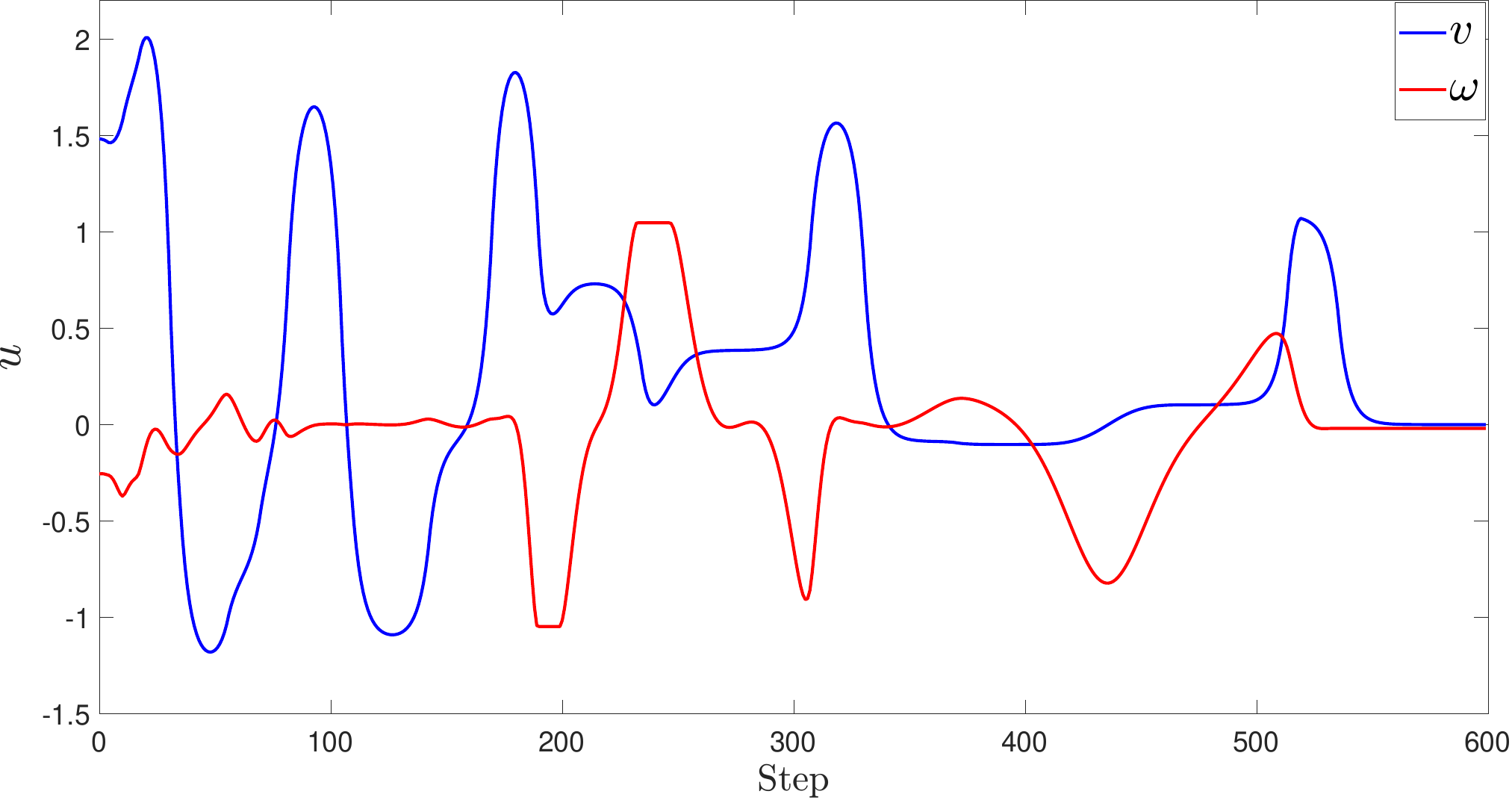}
        \label{fig-3b}
    }
    \hfill
    \subfigure[]{
        \includegraphics[width=0.3\textwidth]{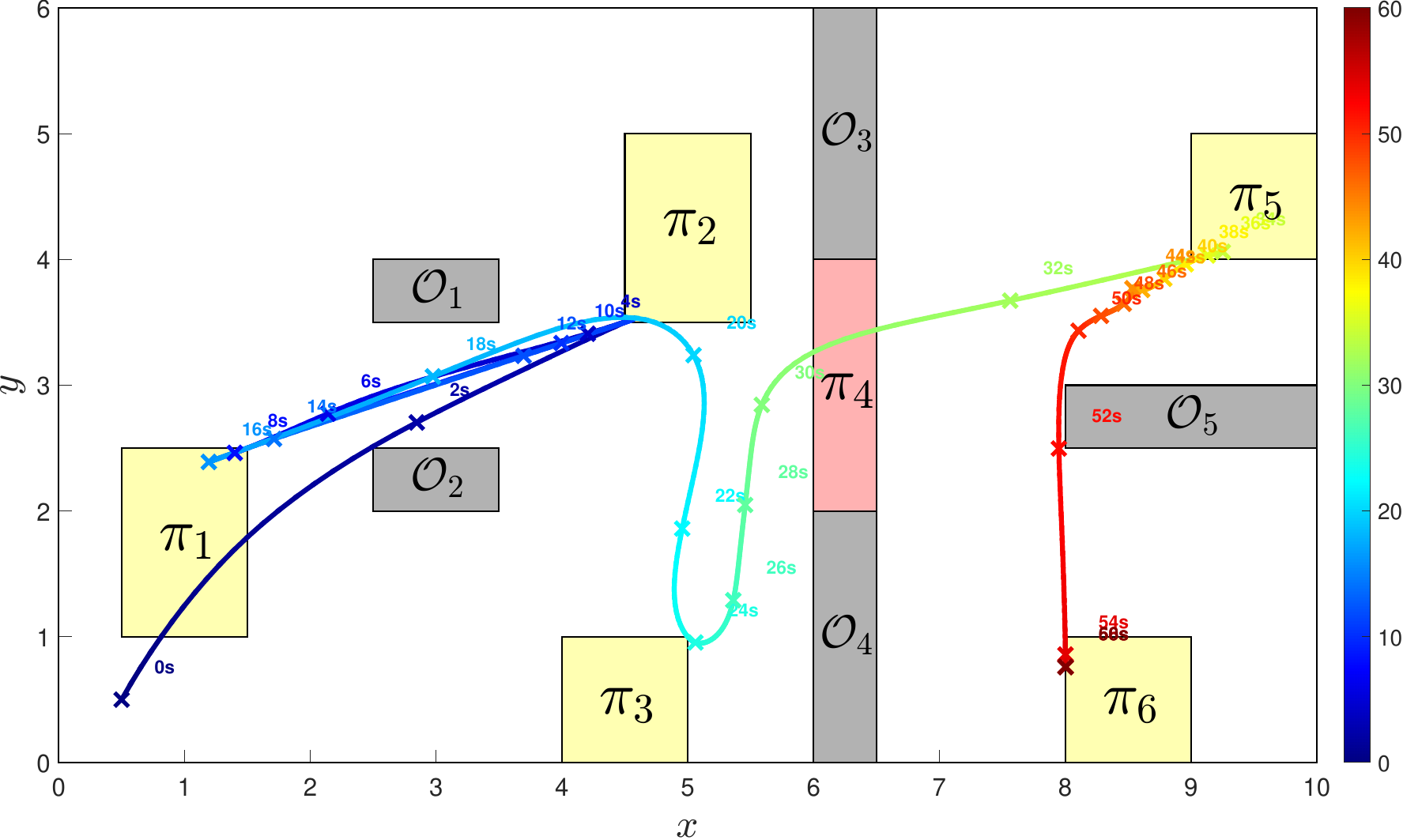}
        \label{fig-3c}
    }
    \caption{Illustration of the simulation results in Scenario 1. Here, $\pi_{r} := \mathcal{R}(\mu_{r}), r \in \{1, \dots, 6 \}$. (a) The results for generating the global waypoints, satisfaction pairs, and safe corridor. (b) The control inputs of the unicycle. (c) The trajectory of the unicycle.}
    \label{fig-3}
\end{figure*}

Through the proposed three-step planning and control approach, the collision-avoidance constraint (5f) and the STL-satisfaction constraint (5g) are reformulated as constraints (22c) and (22d), respectively. Accordingly, the problem (22) is a reformulation of the problem (5). If the problem (22) is feasible, then its solution yields a control sequence whose corresponding position trajectory satisfies $\Psi$ while avoiding obstacles. The optimal control sequence can be computed using nonlinear optimization toolboxes such as CasADi \citep{andersson2019casadi}. Applying this control sequence to system (3) generates the trajectory $\mathfrak{c}$ and the corresponding position trajectory $\mathfrak{p}$.
\section{Numerical Example}
\label{sec-Simulation}

To illustrate the efficacy of the proposed planning and control approach, numerical examples are presented in this section. All computations are carried out via MATLAB R2024a on a laptop equipped with an Intel Core i5-10400 processor and 16 GB of RAM. Consider an autonomous unicycle with the following dynamics from \citep{911382}: 
\begin{equation}
  \label{equ-21}
  \boldsymbol{x}_{k+1} = \boldsymbol{x}_k 
        + \begin{bmatrix}v_k \cos (\boldsymbol{\theta}_k)   \\ v_k \sin (\boldsymbol{\theta}_k) \\ \omega _k \end{bmatrix} \tau,  
\end{equation}
where $\tau := 0.1$ is the sampling period, $\boldsymbol{x} = (\boldsymbol{p}, \boldsymbol{\theta}) \in \mathbb{X}$ is the unicycle state, and $\boldsymbol{u} := (v, \omega) \in \mathbb{U}$ is the control input. $\boldsymbol{p} := (\boldsymbol{p}_{x}, \boldsymbol{p}_{y}) \in \mathbb{P}$ is the position, $\boldsymbol{\theta} \in \Theta$ is the orientation, $v$ is the velocity, and $\omega$ is the angular control input. The input set is $\mathbb{U} := [-4, 4] \times [-\pi/3, \pi/3]$. 

\begin{figure}[htbp]
    \centering
    \includegraphics[width=0.3\textwidth]{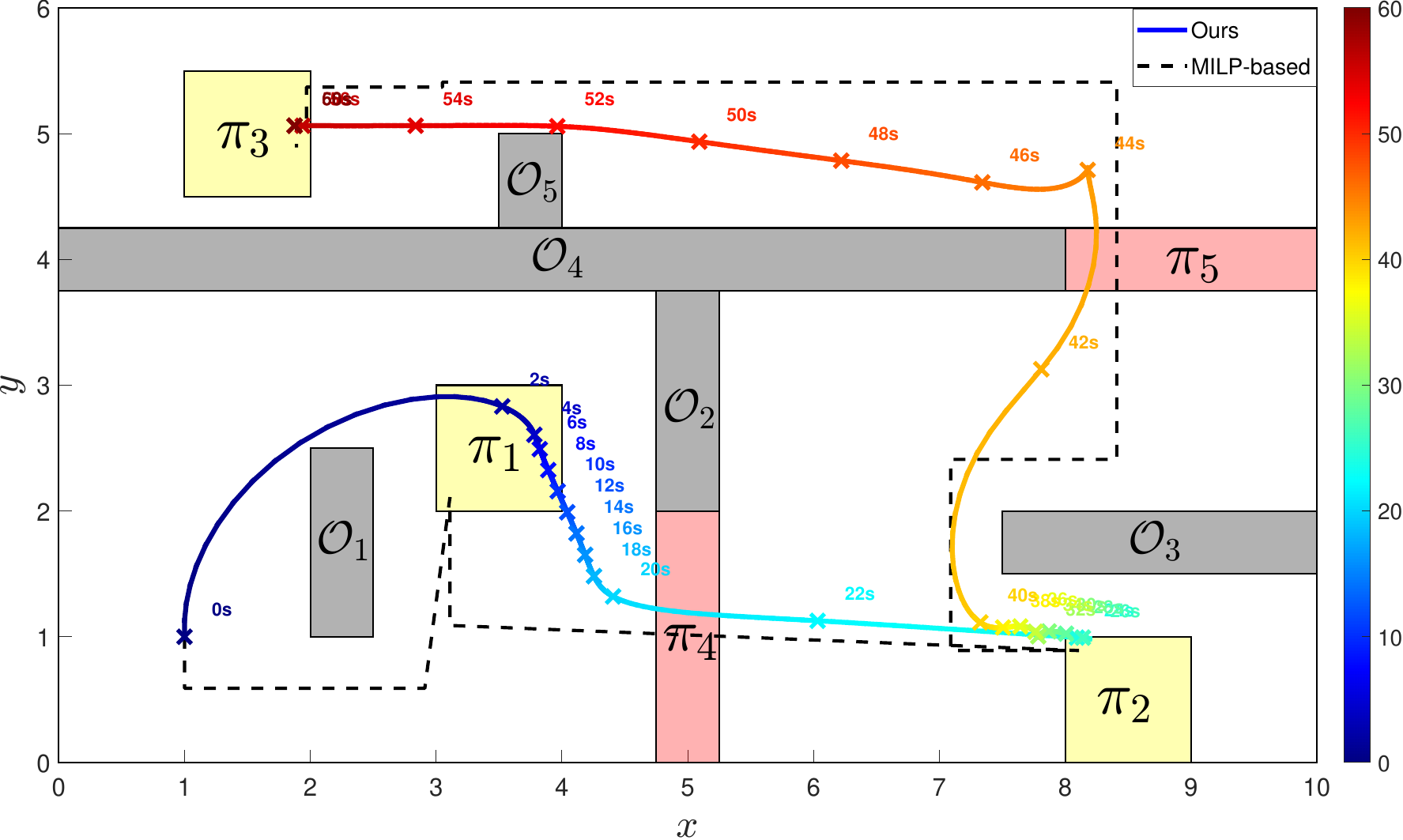}
    \caption{The trajectories of the unicycle in Scenario 2, where $\pi_{r} := \mathcal{R}(\mu_{r}), r \in \{1, \dots, 5 \}$.}
    \label{fig-4}
\end{figure}

\begin{figure}[htbp]
    \centering
    \includegraphics[width=0.3\textwidth]{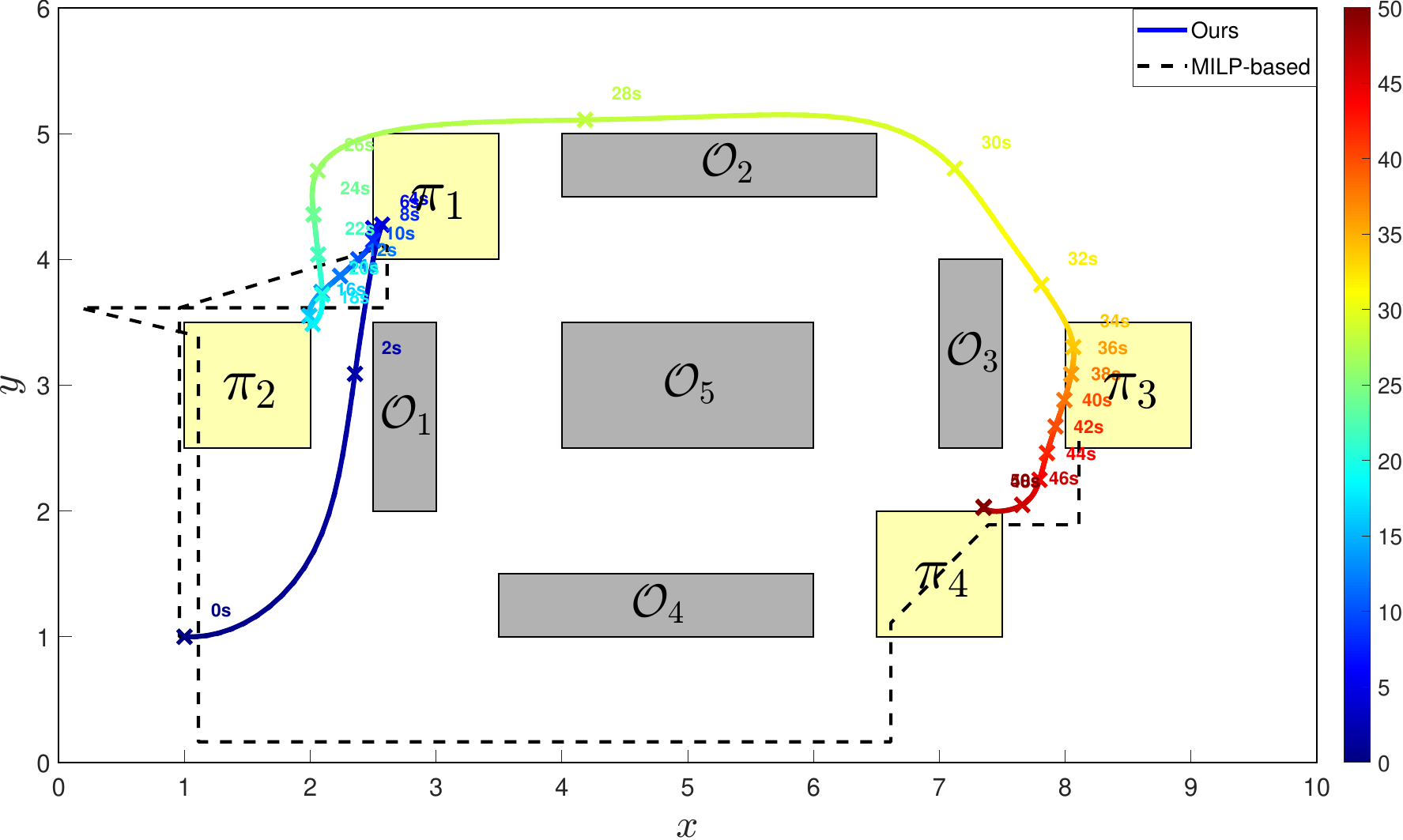}
    \caption{The trajectories of the unicycle in Scenario 3, where $\pi_{r} := \mathcal{R}(\mu_{r}), r \in \{1, \dots, 4 \}$.}
    \label{fig-5}
\end{figure}

For the unicycle \eqref{equ-21},  simulations are performed in the following three scenarios. Our approach is compared with an MILP-based method \citep{sun2022multi}. The corresponding simulation results are shown in Figs. \ref{fig-3}-\ref{fig-5}, and the performance indicators are summarized in Table \ref{tab-1}.
\begin{table}[htbp]
\caption{The comparisons in different scenarios}
\label{tab-1}
\centering
\begin{tabular}{lcc|cc}
\toprule
 & \multicolumn{2}{c}{Ours} & \multicolumn{2}{c}{MILP-based}  \\
\midrule
 & Time(s) & Traj(m) & Time(s) & Traj(m) \\
\midrule
Scenario 1 & 5.30 & 32.15 & - & - \\
Scenario 2 & 3.93 & 20.50 & 1.85 & 23.77 \\
Scenario 3 & 3.44 & 14.97 & 10.66 & 20.50 \\
\bottomrule
\end{tabular}
\end{table}

\textbf{Scenario 1: }In this scenario, $\mathbb{P} := [0, 10] \times [0, 6]$, $\Theta := [-\pi, \pi]$, and there exist $\mathsf{L} := 5$ obstacles. In order to evaluate the efficacy of the proposed approach, the unicycle is initialized at $\boldsymbol{x}_0 := (0.5, 0.5, \pi/3)$ and is required to accomplish the following specifications: $\left. 1 \right)$ within $[0, 10]$ s, $\pi_1$ and $\pi_2$ are to be visited periodically every 10 s; $\left. 2 \right)$ during $[0, 30]$ s, $\pi_4$ should be avoided until $\pi_3$ is reached; $\left. 3 \right)$ between 30 s and 46 s, $\pi_5$ is to be visited and stayed in for 4 s; $\left. 4 \right)$ within [0, 60] s, $\pi_6$ needs to be reached. Hence, the desired task is 
\begin{equation}
  \label{equ-22}
  \begin{aligned}
    \Psi :& = \G_{[0, 10]} \F_{[0, 10]} (\mu_1 \wedge \mu_2) \wedge \neg \mu_4 \U_{[0, 30]} \mu_3 \\
         & \quad \wedge \G_{[30, 46]} \F_{[0, 4]} \mu_5 \wedge \F_{[0, 60]} \mu_6.  
\end{aligned}
\end{equation}

From Section \ref{subsec-DecomTask}, the ordered sequence $\mathbf{S}_{\xi}^{\text{ord}} = (0, 20, 30, 50, 60)$ is derived from \eqref{equ-7}. Hence, $\text{UATI}_1 = [0, 20]$, $\text{UATI}_2 = [20, 30]$, $\text{UATI}_3 = [30, 50]$, and $\text{UATI}_4 = [50, 60]$. $\Psi$ can be decomposed into the following 4 local tasks: $\Psi_1 = \G_{[0, 10]} \F_{[0, 10]} (\mu_1 \wedge \mu_2) \wedge \G_{[0, 20]} \neg \mu_4$, $\Psi_2 =  \G_{[20, 30]} \neg \mu_4  \wedge \F_{[20, 30]} \mu_3$, $\Psi_3 = \G_{[30, 46]} \F_{[0, 4]} \mu_5$, and $\Psi_4 = \F_{[50, 60]} \mu_6$. Then, we apply Algorithms 1-3 to derive the sequence of global waypoints, satisfaction pairs and safe corridor; see Fig. \ref{fig-3a}. Based on the safe corridor and the satisfaction pairs, the optimization problem \eqref{equ-20a} is formulated and solved to derive a sequence of control law and corresponding trajectory; see Fig. \ref{fig-3b} and \ref{fig-3c}. From Fig. \ref{fig-3c}, the STL task \eqref{equ-22} is satisfied by the position trajectory. Note that the position trajectory only reaches the boundaries of $\pi_2$ and $\pi_3$ due to the optimization problem (22). If the cost function is changed, then the robustness of the task satisfaction can be guaranteed. In addition, the MILP-based approach fails to solve the STL task because one of its sub-tasks is the type $\G_{\I_1}\F_{\I_2}\varphi$.

\textbf{Scenario 2: }Scenario 2 is designed to compare the capability of handling non-nested STL formulas. Starting from $\boldsymbol{x}_0 := (1, 1, \pi/2)$, the unicycle is required to accomplish the following STL task
\begin{equation}
  \label{equ-23}
  \Psi := \neg \mu_4 \, \U_{[0, 20]} \mu_1 \wedge \neg \mu_5 \, \U_{[20, 40]} \mu_2 \wedge \F_{[0, 60]} \mu_3.
\end{equation}
From Fig. 4, both approaches successfully generate the trajectories that satisfy the above STL task. However, from Table 1, the proposed approach demonstrates a clear advantage over the method in \citep{sun2022multi} in terms of trajectory length, as it yields a shorter position trajectory that satisfies the STL task (25). Although our method requires more computation time, it is because we simultaneously address both the planning and control problems, whereas the MILP-based method only plans a feasible position trajectory.

\textbf{Scenario 3: }Scenario 3 is designed to evaluate the performance in handling multiple sub-tasks of the type $\F_{\I_1} \G_{\I_2} \varphi$. Starting from $\boldsymbol{x}_0 := (1, 1, 0)$, the unicycle is required to accomplish the following STL task
\begin{equation}
  \begin{aligned}
  \label{equ-24}
  \Psi : & = \F_{[0, 10]} \G_{[0, 5]} \mu_1 \wedge \F_{[0, 25]} \mu_2 \wedge \\
          & \ \quad \F_{[30, 40]} \G_{[0, 5]} \mu_3 \wedge \F_{[30, 50]} \mu_4. 
  \end{aligned}
\end{equation}
From Fig. 5, both approaches are able to generate the trajectories that satisfy the above STL task. However, from Table 1, we can see the advantages of the proposed approach over the one in \citep{sun2022multi} in terms of the computation time and trajectory length. To be specific, with the proposed approach, it takes less computation time to derive a shorter position trajectory to satisfy the STL task (26). In addition, the MILP-based method is to plan a feasible position trajectory only, while both planning and control problems are addressed in this work.
\section{Conclusion}
\label{sec-Conclusion}

In this paper, we proposed a planning and control approach to accomplish the STL task. First, the STL task was decomposed into finite local tasks, each of which can be satisfied by the sequence of local waypoints. The satisfaction pair sets with respect to each local task were derived. Then, a local-to-global strategy was applied to derive the sequence of global waypoints, which was used to construct the safe corridor. Finally, by leveraging the safe corridor and the satisfaction pair sets, an optimization problem was formulated and solved to derive an optimal position trajectory that satisfies the STL task. Future work will focus on the extension of the proposed approach to the multi-robot systems. 


\bibliography{ifacconf}             

\end{document}